# LIGHTWEIGHT HIERARCHICAL MODEL FOR HWSNET

Tapalina Bhattasali[1], Rituparna Chaki[2]

[1]Techno India College of Technology
tapolinab@gmail.com
[2]West Bengal University of Technology
rituchaki@gmail.com

***ABSTRACT***

*Heterogeneous wireless sensor networks (HWSNET) are more suitable for real life applications as compared to the homogeneous counterpart. Security of HWSNET becomes a very important issue with the rapid development of HWSNET. Intrusion detection system is one of the major and efficient defensive methods against attacks in HWSNET. Because of different constraints of sensor networks, security solutions have to be designed with limited usage of computation and resources. A particularly devastating attack is the sleep deprivation attack. Here a malicious node forces legitimate nodes to waste their energy by resisting the sensor nodes from going into low power sleep mode. The target of this attack is to maximize the power consumption of the affected node, thereby decreasing its battery life. Existing works on sleep deprivation attack have mainly focused on mitigation using MAC based protocols, such as S-MAC (sensor MAC), T-MAC (timeout MAC), B-MAC (Berkley MAC), G-MAC ( gateway MAC). In this article, a brief review of some of the recent intrusion detection systems in wireless sensor network environment is presented. Finally, a framework of cluster based layered countermeasure for Insomnia Detection has been proposed for heterogeneous wireless sensor network (HWSNET) to efficiently detect sleep deprivation attack. Simulation results on MATLAB exhibit the effectiveness of the proposed model.*

## 1. INTRODUCTION

Wireless sensor network (WSNET) consists of low-cost, resource limited sensor nodes to sense data and to transmit it to distant sink node that acts as gateway to another network or access point for human interface. In reality, homogeneous sensor networks hardly exist. In heterogeneous sensor network, a large number of inexpensive nodes perform sensing, while a few nodes having comparatively more energy perform other tasks such as data filtering, transport. This leads to the research on heterogeneous wireless sensor network (HWSNET) where different types of nodes are considered to prolong the life-time and reliability of the network. HWSNET is a rapidly growing area in the real life applications like traffic, environment monitoring, healthcare, military applications, home automation, disaster monitoring.

WSNET suffers from several drawbacks like limited energy, computation, and storage capabilities of sensor nodes. It is more vulnerable to various attacks due to its deployment in hostile environment where it is very difficult to change or recharge the batteries. A particularly devastating attack is the sleep deprivation attack, where a malicious node forces legitimate nodes to waste their energy by resisting the sensor nodes from going into low power sleep mode. In 1999, Stajano and Anderson [10] first mentioned denial of sleep attack as "sleep deprivation torture". The target of this attack is to maximize the power consumption of the affected node, thereby decreasing its battery life. This attack can nullify any energy savings obtained by allowing sensor nodes to enter into sleep mode. Sleep deprivation attack comes also in the form of sending traffic that causes a sleeping node to wake-up. This attack can nullify any energy savings obtained by allowing sensor nodes to enter into sleep mode. This type of attack has the potential to lessen the lifetime of sensor nodes. It imposes such a large amount of energy consumption upon the limited power sensor nodes that they stop working. This gives rise to denial of service by violating wireless network connectivity. The attacker launches a sleep deprivation attack by interacting with the victim in a manner that appears to be legitimate. As a consequence, this attack is difficult to detect because it is assumed to be carried out through the use of seemingly innocent interactions.

The typical IDS used in MANETs fail to provide power efficiency. Thus security solutions in WSNET have to be designed with efficient resource utilization, especially power. Existing works on sleep deprivation attack have mainly focused on mitigation using MAC based protocols, such as S-MAC (sensor MAC), T-MAC (timeout MAC), B-MAC (Berkley MAC), G-MAC (gateway MAC). But it may increase the overhead of security solution in terms of complexity and cost. Therefore the requirement is to found the alternative of MAC based protocols for detecting sleep deprivation attack.

The rest of this article is organized as follows: section 2 introduces the intrusion detection system and related technologies for WSNET. In section 3, an insight into ongoing research activities is presented. Section 4 consists of an outline of the proposed model. Chapter 5 illustrates the experimental results for performance analysis. It is followed by a conclusion in Section 6.

## 2. Intrusion Detection System for WSNET

An intrusion can be defined as a set of actions that attempt to compromise the integrity, confidentiality or availability of a resource. In general there are two major kinds of detection techniques, (i) anomaly detection, (ii) misuse detection. Anomaly detection technique detects unusual deviations from the normal behavior. The determination of threshold values for normal behavior becomes very important in this type of detection. This has high probability of false alarm rate because anomaly detection generally uses a defined model of normal behavior; a packet is determined to be abnormal by the system when the current behavior varies from the model of normal behavior. As a result, anomaly detection often classifies normal behavior as abnormal, leading to the problem of erroneous classification. However, it seldom marks an abnormal behavior as normal. Thus, detection system that can adaptively change the normal model is preferable. Anomaly detection properties that must be satisfied are – accuracy, responsiveness, resource usage, robustness. Misuse detection technique (signature based) compares patterns (signatures) of well-known attacks stored in profiles with the current scenario. This technique fails to detect unknown attack patterns.

Intrusion detection system (IDS) detects security violation on a system by monitoring and analyzing network activities, and raises an alarm when an intrusion occurs. The constraints inherent to sensor network, such as sparse resource and limited battery life, impose a restriction on how the detection tasks are performed. The nature of sensor networks necessitate that the efficient IDS must fulfill the following requirements:

- ✓ IDS must be able to detect the anomalies with considerable accuracy.
- ✓ Detection and responses to anomalies must be within acceptable time period.

- ✓ It must be capable of isolating intruders successfully in WSNET.
- ✓ IDS must be lightweight consuming less energy (minimal computing and battery) to extend WSNET life cycle.

In wireless sensor networks, IDS topology can be classified as follows:

- ➢ **Distributed approach:** Intrusion detection load is divided among the sensor nodes, which may collaborate with each other to form a global intrusion detection mechanism. This architecture is more suitable for flat wireless sensor networks.
- ➢ **Hierarchical approach:** This architecture has been proposed for multilayered wireless sensor network. In this approach, network is divided into clusters where cluster-heads aggregate data collected from the member nodes. At the same time all cluster-heads can cooperate with central base station to form a global IDS.

Most of the existing intrusion detection techniques have not met the requirements for practical deployment in wireless sensor network.

## 3. RELATED WORKS

This section presents a category-wise report of on-going research activities. IDS can be implemented using various techniques. Intrusion detection for WSN is an emerging field of research.

In [1], a semantic based intrusion detection framework is proposed for WSN by using multi-agent and semantic based techniques, where security ontology is constructed according to the features of WSN to represent the formal semantics for intrusion detection. This distributed technique is based on cooperative mechanism. IDS framework includes the following layers: (1) network layer refers to the network topology of the WSN; (2) semantic layer refers to security ontology (formal semantics for WSN activities); (3) model layer refers to intrusion detection model (collection of rules for intrusion detection) for single sensor node. The model determines the behaviors of sensor nodes; (4) cooperative layer refers to the policy that how sensor nodes cooperate with each other for detecting intrusion. In this mechanism, each selected rule of security ontology is mapped to sensing data collected from common sensor nodes to detect anomaly. If result of intrusion detection by an agent node (monitor node that is only equipped with intrusion detection model) is undecidable then data is forwarded to a neighboring agent node.

In [2], an energy efficient learning solution for IDS in WSN has been proposed. This schema is based on the concept of stochastic learning automata on packet sampling mechanism. Simple Learning Automata based ID (S-LAID) functions in a distributed manner with each node functioning independently without any knowledge about the adjacent nodes. System Budget (total sampling budget of a single node) is analogous to the amount of energy that the node can spend on intrusion detection during its lifetime and the balance budget of the system is analogous to the residual sampling energy of the system. Exhaustion of sampling budget of a node implies that no more energy can be spent on intrusion detection tasks. In S-LAID, each node continuously samples its interface at a minimum sampling budget. If malicious packets are found and the detection rate is more than the penalty threshold, then the sampling rate is increased by penalization function. When the detection rate is less than the penalty threshold, the sampling rate is decreased by reward function. The rate control algorithm is used to control the increment in the sampling rate. System checks whether it is profitable to increase the sampling budget to make sure that it works in energy efficient manner.

In [3], a location-aware, trust-based detection and isolation mechanism of compromised nodes in wireless sensor network is proposed. In this technique, probabilistic model is used to define trust and reputation. After deployment, each node periodically broadcasts one-hop hello packets to discover its neighbors. If any node is verified to be authentic, it is recorded in its neighbors list and its trust value is initialized. Then a secure cluster formation algorithm is executed. The maintenance phase involves updating reputation, trust, and confidence metrics according to the modeling parameters. During this phase the nodes monitor the traffic coming in and out of their neighbors. The purpose of revocation is to remove dishonest nodes from the network. Each node periodically checks its trust table for detecting lowest trust metric. If the least trusted node also has a confidence value above a predetermined threshold then that node is blacklisted and its node id is broadcasted as being dishonest. If there is a tie then one of the least trusted nodes is randomly selected for broadcasting its id and it has to be decided whether this node need to be blacklisted or packet need to be dropped.

In [4], a method using isolation tables is proposed to isolate malicious nodes by avoiding consumption of unnecessary energy by IDS (ITIDS).This hierarchical structure of IDS based on cluster network can detect serious attacks such as hello flooding, denial of service (DoS), denial of sleep, sinkhole and wormhole attack. In this model, load of PCH (primary cluster head) is distributed among number of SCHs (secondary cluster head) that can directly receive sensing data from each MN (member node) of MGs (Monitor Group).Base station receives sensing data and isolation tables from PCH whose responsibility is to gather sensing data and integrate isolation tables from SCH, which is selected randomly. SCHs calculate trust values to find malicious MNs (member nodes), isolate and record malicious nodes in its isolation table. MN can determine whether the PCH is intruder or not. In this mechanism, malicious nodes can be detected by considering remaining energy and trust values of sensor nodes.

In [5], a lightweight ranger based IDS (RIDS) has been proposed. It combines the ranger method to reduce energy consumption and the isolation tables to avoid detecting anomaly repeatedly. This lightweight IDS model relates ontology concept mechanism about anomaly detection. In this technique, rough set theory (RST) is used for preprocessing of packets and anomaly models will be trained by support vector machine (SVM). Relationship between sensor node data and ontology concept is compared to detect attack. The ranger loads detection models for detecting anomaly and the error information will be recorded in isolation table of the base station. Different attack type modules are selected depending on environment of sensor network to establish lightweight IDS. Ontology is applied to construct relationship between nodes of WSN. The IDS calculates relationship of whole WSN to define relationship threshold.

In [6], a hierarchical overlay design (HOD) based intrusion detection system is proposed, using policy based detection mechanism. This model follows core defense strategy where cluster-head is the centre point to defend intruder and concentrates on saving the power of sensor nodes by distributing the responsibility of intrusion detection to three layer nodes. Except leaf nodes, all intermediate nodes have high energy. Each area of sensor nodes is divided into hexagonal regions where sensor nodes are monitored by cluster node which is in turn monitored by regional node that is controlled and monitored by the base station. The HOD based IDS uses hybrid approach (misuse and anomaly) of intrusion detection mechanisms. The components of IDS for policy based management include (1) Base policy decision point (BPDP) is the controlling component of the architecture. It implements intrusion rules generated by the intrusion detection tool (IDT) from receiving events, evaluating anomaly conditions.(2) Local policy agents (LPAs) and regional policy agents (RPAs) act as policy decision modules (PDMs). LPA reduces management bandwidth and computational overhead from leaf level sensor nodes to improve network performance and intrusion detection efficiency. An RPA can manage multiple LPAs and BPDP manages and controls all the RPAs.(3) Policy Enforcement Points (PEP) are low level sensor nodes. In this technique, if regional node fails then base station can select one of its neighbor nodes dynamically according to some predefined rule in BPDP. Then BPDP needs to supply the policy, rules, or signatures of failed node to the selected new neighbor regional node. In the same way RPA has the only responsibility to select appropriate neighbor LPA. As base station is much more powerful node with large storage; all the signatures, anomaly detection rules or policies are stored primarily as backup in base station.

In [8], a hierarchical model (three layer architecture) is proposed based on weighted trust evaluation (WTE) to detect malicious nodes by monitoring its reported data. It is a three layer architecture which consists of three types of sensor nodes: (1) Low-power, limited functionality sensor nodes (SN) that do not support multi-hop routing capability to its neighbor.(2) High-power, trustful forwarding nodes (FN) that forward data obtained form sensor nodes to upper layer and offers multi-hop routing capability to SNs or other FNs.(3) Access points (AP) or base stations (BS) that route data between wireless networks and the wired infrastructure and considered to be trustful. This model considers only HELLO flooding attack, Sybil attack, blackhole attack, wormhole attack, DDoS attack. In this mechanism, FN collects all information provided by SNs and calculates an aggregation result (E) by using the weight ($W_n$ ranging from 0 to 1) assigned to each SN and sensor node's output that may be true(alarm) or false(no alarm). If a sensor node is compromised and frequently sends its report inconsistent with the final decision, its weight is likely to be decreased. Based on updated weights, the forwarding node is able to detect a node as a malicious node if its weight is lower than a specific threshold.

In [9], a dynamic model of intrusion detection (DIDS) has been proposed for WSN. This is a hierarchical model of IDS based on clustered network to battle the low energy. The components of IDS framework are an event monitor module for monitoring event data in its radio range, rule record base for storing rules to detect unauthorized access, alert for producing signal if intrusion is detected by IDS, a misuse detection module for generating alert for known attacks, anomaly detection module to generate alert for any generated anomaly. The IDS is pre-installed in every sensor node, and gets activated at specific times. If a node is found to consume more energy after activation of IDS, a cluster reconfiguration process is proposed for energy optimization. When number of detected intruders per unit time in a cluster is more than the threshold value, then the procedure upgrades the core defense and boundary defense mechanism to distributed defense mechanism (selecting agent node by voting algorithm) which has stronger detection capability. It can use distributed defense which has the advantage of detecting multiple intruders, albeit, with an increased rate of energy consumption with increase in cluster size.

## 4. PROPOSED MODEL

In this section, a lightweight, hierarchical model is proposed for heterogeneous wireless sensor network (HWSNET) to detect insomnia of sensor nodes affected by sleep deprivation attack. It uses cluster based mechanism in an energy efficient manner to build a five layer hierarchical network to enhance network scalability and lifetime. The low energy constraints of WSNET necessitate the use of a hierarchical model. Here sensor network is divided into clusters which are again partitioned into sectors. Partitioning the sensor field can conserve communication bandwidth and avoids redundant exchange of messages among sensor nodes. It can prolong the battery life of the individual sensors and the network lifetime and can reduce the rate of energy consumption. In this model, energy efficiency can be achieved by keeping a minimal number of sensors active. A dynamic model is designed here to overcome sudden death of IDS enabled sensor nodes which are responsible for all detection tasks, due to exhaustion of power. The proposed model uses anomaly detection technique in such a way so that phantom intrusion detection can be avoided.

### 4.1 Assumptions

Assumptions are necessary to define both the boundaries of this research and the scope of the problem. The following assumptions are taken in order to design the proposed model.

- Different states of a sensor node:

  **NEW→MEMBER→ SUSPECTED→MALICIOUS→ISOLATED**

  ↓ ↕ ↓

  **GENUINE → DEAD**
- Every node has a unique id (geographical position vector) in the network.
- Each member node has member-id and authentic wake-up coin.
- A protocol is used to assign a secure optimal wakeup and sleep schedule for the sensor nodes.
- Sink node is honest gateway to another network or access point.
- Threshold values are pre-calculated and set for the entire network.
- Timestamp must be included in each message during transmission through network.
- SM may be more than one within a sector.
- SN selects CC, SM, FSH and CC selects SC.

## 4.2 System Model

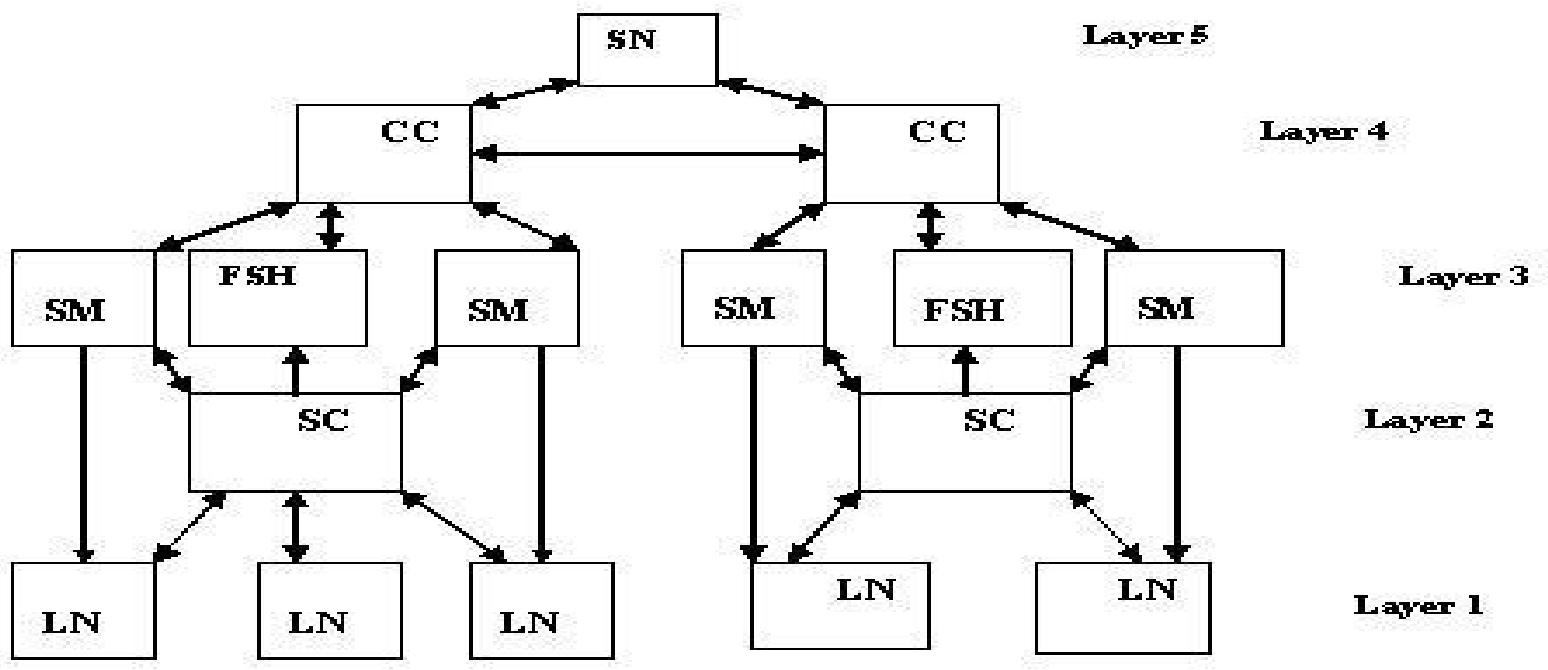

**Figure 1 Layered Model**

Figure 1 describes the main building block of the system model. Here SN→SINK NODE; CC→CLUSTER COORDINATOR; SM→SECTOR MONITOR; FSH→FORWARDING SECTOR HEAD; SC→SECTOR COORDINATOR**;** LN→ LEAF NODE;

## 4.3 Layers of System Model

The five layers of proposed model are described below-

**Layer 1**: In this lowest layer leaf nodes sense environmental data and send it to its immediate next higher layer i.e. layer 2. Layer 1 has no anomaly detection capacity.

**Layer 2**: This layer includes sector coordinator (SC) of each sector. Sector coordinator maintains membership list [] → {node-id, member-id, node status, validity} of all leaf nodes within a sector, normal profile [] (tuple space that consists of sensor node's attribute) and knowledge base [] (system parameters, application requirements), reputation list [] → {node-id, member-id, scount , ncount, trust, na, belief, reput}, suspected list [] → {node-id, member-id, na, monitor, trust ,scount ,Tot, suspected}.

**Layer 3**: This layer includes sector monitor (SM) and forwarding sector head (FSH). Sector monitor maintains suspected list [], normal profile [], knowledge base [], reputation list [], quarantine list [] → {node-id, member-id, na, monitor, compromised, trust, scount, malicious}. FSH (nearest neighbor of cluster coordinator) acts as router that inserts valid packet details to forwarding table [] → {node-id, member-id, na, packet-id, node-info, next hop, timestamp}.

**Layer 4**: This layer constitutes the cluster coordinator (CC) which controls SM and FSH of each sector within a cluster. It inserts valid packets details to valid list [] → { node-id , member-id, na, reputation} and forwards data to the sink node. Cluster coordinators (CC) can cooperate with each other to form global IDS.CC contains backup copy of its own cluster.

**Layer 5**: The topmost layer is the sink node that collects data from lower layer and it acts as a gateway between sensor network and other networks or acts as access point. SN contains backup copies of all clusters.

## 4.4 Data Definition

*Definition 1:* ***Leaf Node*** **LN–** A node N is defined to be a Leaf Node if $Child_N$ { }= {∅} AND $Parent_N$ { } ≠ {∅}. Its detection power (DP) ←0 AND priority level← MIN_PRIORITY(5).

*Definition 2:* ***Sector Coordinator SC –*** A node N is defined to be a Sector Coordinator if $Child_N$ { } ≠ {∅} AND $Parent_N$ { } ≠ {∅} AND $hop_distance_N$ = min{hop_ distance[FNi] from $CC_k$}, $Rem_eng_N$ = MAX_ENG {FNE[i]}, where FNE[i] → remaining energy of follower nodes. Its priority level ← 4.

*Definition 3:* ***Sector Monitor SM*** - A node N is defined to be a Sector Monitor if $Child_N$ { } ≠ {∅} AND $Parent_N$ { } ≠ {∅} AND $Rem_eng_N$ = MAX_ENG {LNE[i]}, where LNE[i] → remaining energy of all leader nodes excluding cluster coordinator, $DP_N$ = MAX_DETECT {N[i]}, where N∉ {$CC_k$, SN} AND $DP_N$→ power required for intrusion detection. Its priority level←3.

*Definition 4:* ***Forwarding Sector Head FSH*** - A node N is defined to be a Forwarding Sector Head, $Child_N${ } ≠ {∅} AND $Parent_N$ { } ≠ {∅} AND $hop_distance_N$ = min{hop_ distance[i] from $CC_k$},where N ∉$CC_k$. Its detection power (DP) ←0 AND priority level ←3.

*Definition 5:* ***Cluster Coordinator CC*** - A node N is defined to be a Cluster Coordinator, if $Child_N${ } ≠ {∅} AND $Parent_N$ { } ≠ {∅} AND $hop_distance_N$ = min{hop_ distance [LNi] from SN}, where N ∉SN , $Rem_eng_N$ = MAX_ENG {N[i]}. Its priority level←2.

*Definition 6:* ***Sink Node SN*** - A node N is defined to be a Sink Node if $Child_N$ { } ≠ {∅} AND $Parent_N$ { } = {∅}. Its priority level←MAX_PRIORITY (1).

## 4.5 Insomnia Detection Model

The entire heterogeneous sensor field is divided into overlapping or disjoint clusters like $C_k$, for k ∈ {1,..,r}, r being the number of clusters in the sensor network. Each cluster consists of its member nodes including a cluster coordinator. Let $mem_1$, $mem_2$, ....,$mem_n$ be the members of a cluster $C_k$, and n is the number of members within a cluster excluding CC. Clusters are partitioned into non-overlapping sectors like $S_j$, for j∈{1,…,m},where m is the number of sectors within a cluster. Three types of sensor nodes are assumed in this five layered model: (i) leader nodes or LDN (in layer 3 and 4) (ii) follower nodes or FN (in layer 1 and 2) and (iii) sink node or SN (in layer 5). Leader nodes can be equipped with EXIDS (extended IDS), but only the node designated as sector monitor can activate it. Cluster coordinator (CC) and sink nodes (SN) are also using EXIDS for detecting intrusion during emergency. SN acts as controller and CC acts as assistant controller. SIDS (simple IDS) can be loaded in all follower nodes, but can be activated only at sector coordinator of layer 2 for detecting anomaly. Leaf nodes periodically generate data, possibly at different rates, for transfer to the sink node. Local timestamp is inserted in each packet. Sector coordinator collects sensing data within allotted time slot of each leaf node in a sector. A schedule assigns one or more time slots to each node. At time slot $T^i$ the sector coordinator receives an acquisition vector av[i] and computes the probability of anomaly. Sector coordinator (SC) monitors the sensor nodes for detecting anomaly by SIDS. Suspected nodes are penalized and legitimate nodes are rewarded. Forwarding sector head (FSH) forwards valid packets to CC. Sector monitor (SM) decides whether a suspected node is malicious or not. EXIDS has the responsibility to declare the suspected node as malicious and to drop fake or corrupted packets. To avoid phantom intrusion detection logically, suspected nodes get chance to increase their reputation by SM, if it is not decided as malicious. Intruder or malicious nodes are isolated in quarantine list; so that no intrusion occurs through these nodes.

When detection power reaches to minimum threshold, detection capacity is automatically disabled. Reconfiguration procedure takes place dynamically if any node found to be suspected i.e. energy consumption rate greater than normal consumption rate. Each of leader and follower nodes must be included within a cluster. If any node is under more than one CC, then its membership to the cluster is determined by checking that which CC can provide the strongest signal to that sensor node. If there is a tie, it is broken randomly. Anomaly is detected by SC. But there is a possibility of false positive or false negative. If any genuine node is suspected by SC (false positive), SM can detect it and takes final decision. If any compromised node is treated as genuine and forwarded to FSH (false negative), CC can detect it. SN selects cluster coordinator on the basis of leader node's maximum current residual energy and minimum distance among all other nodes and coverage area of CC is considered as cluster. Then CC selects SC on the basis of follower node's maximum current residual energy and minimum distance among all follower nodes. The coverage area of SC is considered as sector. Sector monitor within a sector is selected on the basis of its maximum current residual energy and detection power. FSH is selected on the basis of minimum distance from CC.

## 4.6 Procedural Steps

The proposed model can work according to the following steps:

- **Initialization Phase:** Sensor nodes are deployed in the sensor field during this phase. A unique identification number consisting of the geographical position vectors is assigned to each new node. The sink node searches for its neighbors to acquire energy details of all nodes after broadcasting advertised message.
- **Cluster Coordinator Selection and Cluster Formation Phase:** Cluster coordinator is selected among all leader nodes and its coverage area is considered as cluster. The Cluster-head details are broadcasted to all its neighbors. The neighbor nodes collect advertised messages during a given time interval and send a join message to nearest cluster coordinator for all nodes within the range of any specific cluster coordinator. Intersection of domains of two clusters may or may not be NULL.
- **Sector Coordinator Selection and Sector Formation Phase:** Sector coordinator is selected among all follower nodes and its detailed information along with node-id is broadcasted to all of its neighbors. Its coverage area is considered as sector. Intersection of domains of two sectors must be NULL. Sector monitors and forwarding sector heads are selected for each sector.
- **IDS Activation Phase:** Activate IDS preinstalled in cluster coordinators, sector monitors and sector coordinators.
- **Reconfiguration Phase:** When cluster coordinator or sector coordinator's behavior deviates from normal, reconfiguration procedure takes place.

- **Data Transfer and Intrusion Detection Phase:** After sector coordinator selection is done each follower node (leaf node) sends data to the sector coordinator that transfers genuine packet to its cluster coordinator through forwarding sector head. Cluster coordinator collects valid data from all sectors within its coverage area and then forwards aggregated packets to sink node.

## 4.7 Detection System of Proposed Model

The proposed model contains detection system consisting of two layers: i) SIDS; ii) EXIDS.

**SIDS: First Layer of IDS**

Simple Intrusion Detection system can only capable of detecting anomaly and generating alert for suspected node. Although it consumes less power in detection, it may give rise to phantom detection.

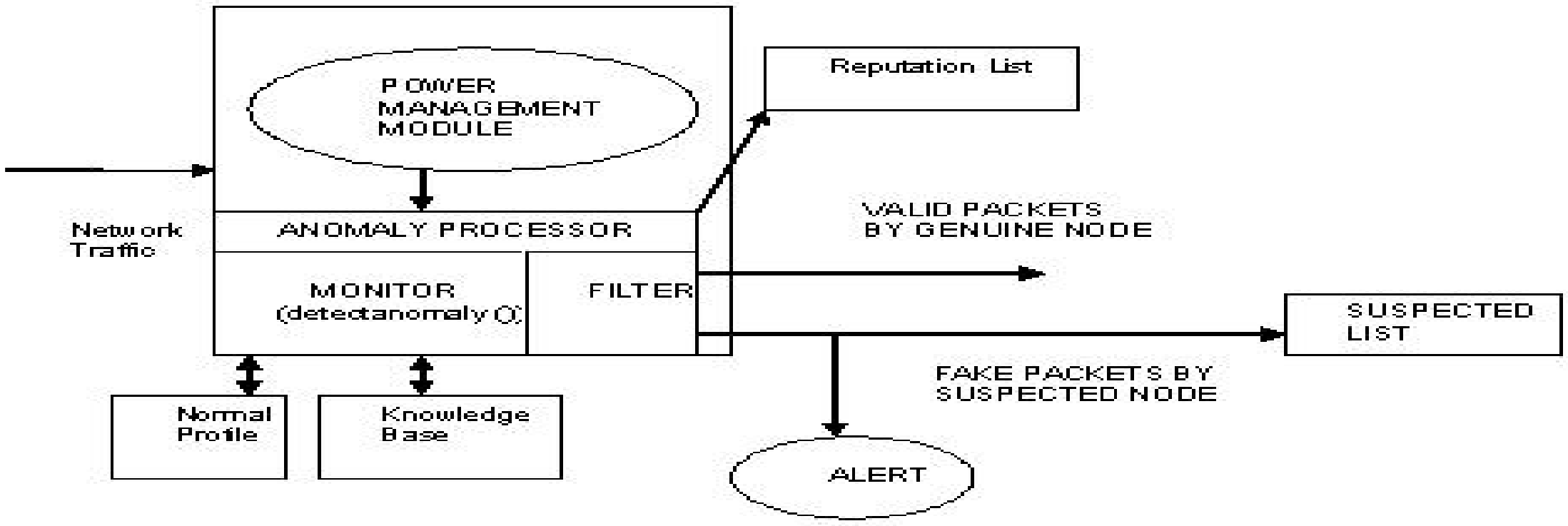

**Figure 2 SIDS Model**

**EXIDS: Second Layer of IDS**

Extended Intrusion Detection system can take final decision of intrusion detection. Features of SIDS are extended here. It consumes more power in detection, but it can reduce phantom intrusion detection.

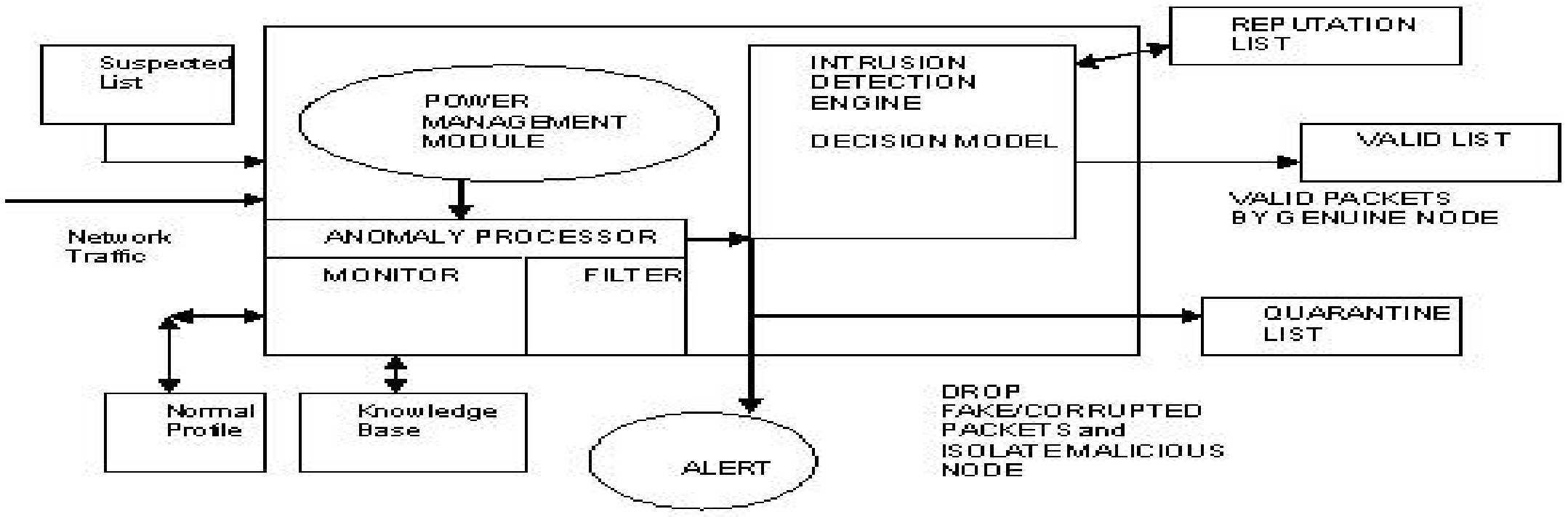

**Figure 3 EXIDS Model**

Table 1. Terminologies used in Proposed Model

| Term | Meaning |
|---|---|
| $p^i_{jk}$ | Packet sent by $i^{th}$ leaf node of sector j of cluster k |
| $LN^i_{jk}$ | $i^{th}$ leaf node of sector j of cluster k |
| $T^i$ | Time-slot allocated to $i^{th}$ leaf node |
| insomnia | Result of anomaly detection |
| $L^i_{jk}$ | Standard battery lifetimeof $i^{th}$ leaf node of sector j of cluster k |
| $PW^i_{jk}$ | Initial battery power of $i^{th}$ leaf node of sector j of cluster k |
| $RE^i_{jk}$ | Residual energy of $i^{th}$ node of sector j of cluster k |
| $CRLT^i_{jk}$ | Calculated remaining lifetimeof $i^{th}$ node of sector j of cluster k |
| $LRE^i_{jk}$ | Last recorded energy of $i^{th}$ node of sector j of cluster k |
| $Tot^i_{jk}$ | Total number of packets sent by $i^{th}$ leaf node of sector j of cluster k |
| NP[] | Normal profile |
| KB[] | Knowledge base |
| $R_{suspected}$ | Rate of energy consumption for suspected node |
| Truedetect | Count the number of times system detects true intrusion |
| Phantomdetect | Count the number of times system detects false intrusion |

## System Parameters

$PW_{slp}$ → power required in sleep mode; $PW_{tr}$ → power required during transmission; $PW_{idle}$ →power required in idle mode; $PW_{wake}$→ power required in wakeup mode; $PW_{comp}$ →power required during computation; $PW_{sensing}$→ power required during sensing data; $T_{idle}$→ time spent in idle mode; $T_{slp}$→ time spent in sleep mode; $T_{tr}$→time spent during transmission; $T_{comp}$ →time spent in computation; $T_{wake}$→ wakeup duration; $T_{sensing}$ →sensing duration; $NEC_{ijk}$→ normal energy consumption; $TNEC_{ijk}$ → threshold normal energy consumption; $ThL_{ijk\rightarrow}$ threshold lifetime; $ThT_{wk}$→threshold wakeup duration; $ThT_{sl}$→ threshold sleep duration; $Th_{buf}$ →threshold buffer capacity; $AWC^i_{jk}$ →authentic wake up coin value; $T_{scount}$ →threshold of allowable suspected count; $T_{per}$→ threshold of allowable suspected count percent; $T_{reput}$ → threshold reputation; Th →threshold value

## 4.7 Insomnia Detection Procedure

Begin

**Case 1**: **/* Energy consumption rate of any node found to be more compared to preset threshold value of normal energy consumption or calculated lifetime of any node found to be less compared to preset threshold lifetime of the node*/**

If $EC^i_{jk} > TNEC^i_{jk}$ OR $CLT^i_{jk} < THL^i_{jk}$ then
    insomnia ←*1*
Else
    insomnia ←*0*
EndIf

**Case 2**: **/* Allotted wakeup period of any node is greater than predefined threshold wakeup schedule and allotted sleeping period is less than predefined threshold sleeping schedule*/**

If $T_{wake} > ThT_{wk}$ AND $T_{slp} < ThT_{sl}$ OR $T_{slp} = 0$ then

insomnia ←*1*

Else

insomnia ←*0*

EndIf

**Case 3**: **/* Any node sends packets in a time-slot, but that slot is not allocated to that node*/**

If | $T_{slot}{}^{i}{}_{jk}$ -$T^i$ | >0 then // $T_{slot}{}^{i}{}_{jk}$ → data actually transmitted by $LN^i_{jk}$ during this time-slot

insomnia ←*1*

Else

insomnia ←*0*

EndIf

**Case 4**: **/* Residual energy of any node is found to vary more compared to last recorded energy*/**

If $LRE^i_{jk}$>> $RE^i_{jk}$ OR $LRE^i_{jk}$<< $RE^i_{jk}$ then

insomnia ←*1*

Else

insomnia ←*0*

EndIf

**Case 5**: **/*Received packets within a time interval exceeds pre-defined threshold value*/**

If ($Tot^i_{jk}$ / $T^i$ )*100% > $Th_{buf}$ then

insomnia ←*1*

Else

insomnia ←*0*

EndIf

**End Case**

End

## 5. Performance Analysis

In this section, the analysis of proposed model is validated using MATLAB 7.0. Performance has been studied for existing ITIDS and proposed model. In figure 4, the result shows that, number of alive nodes with respect to increasing time in second is more in proposed model. Therefore it can be said that our HWSNET lifetime is better than ITIDS. Because proposed model uses dynamic configuration and cluster is further partitioned into sectors. In figure 5, the result shows that accuracy is comparatively high in proposed model because here sector monitors which have high detection power are used to detect intrusion; whereas in ITIDS low energy member nodes are considered as monitor nodes. In figure 6, energy consumption is compared with respect to the density of sensor nodes with clusterization and sectorization and without clusterization or sectorization. The result shows energy consumption is comparatively less when sensor field is partitioned into clusters and sectors. In figure 7, result shows that packet transmission overhead increases with time. After analyzing performance, it can be said that proposed model can prolong network lifetime, detect intrusion accurately with less delay and consumes less energy to detect sleep deprivation attack.

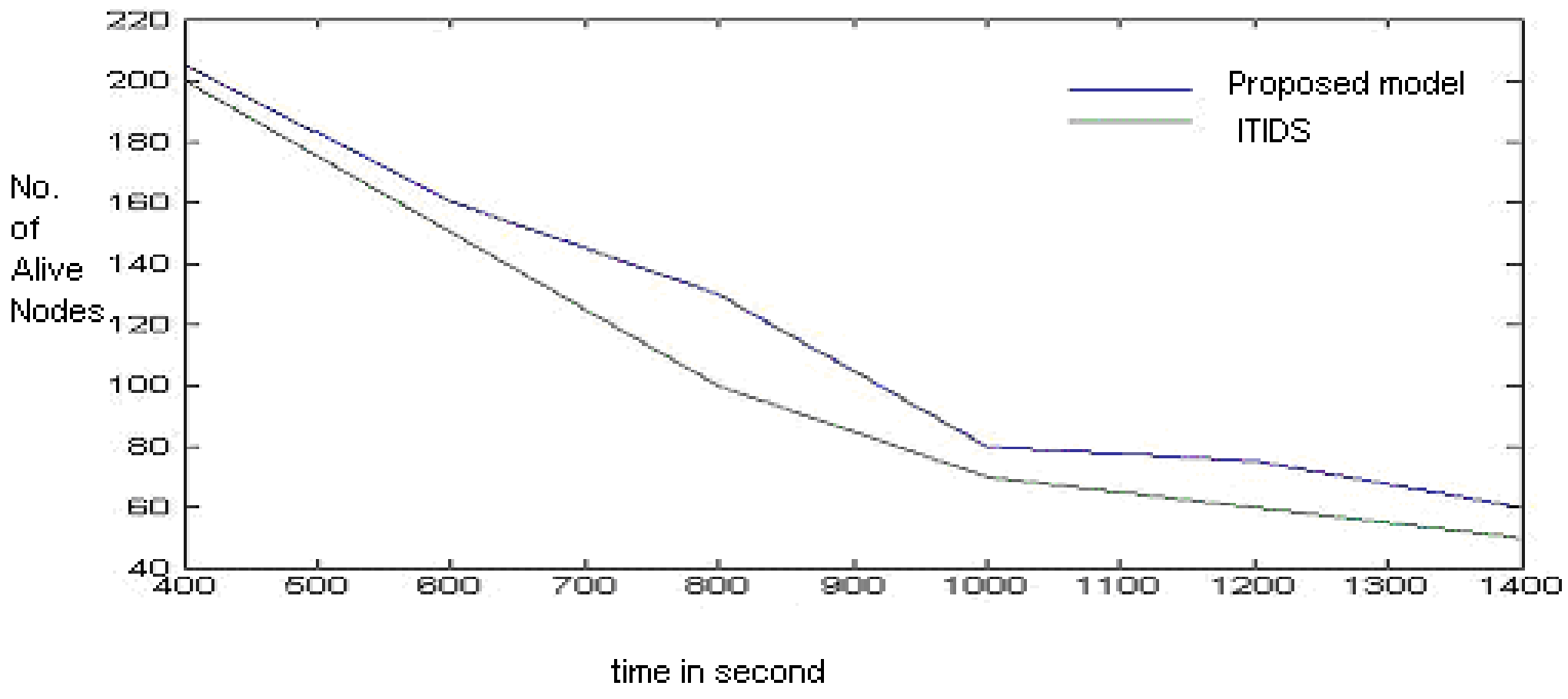

Figure 4 Comparison of Number of Alive Nodes Between ITIDS and Proposed Model

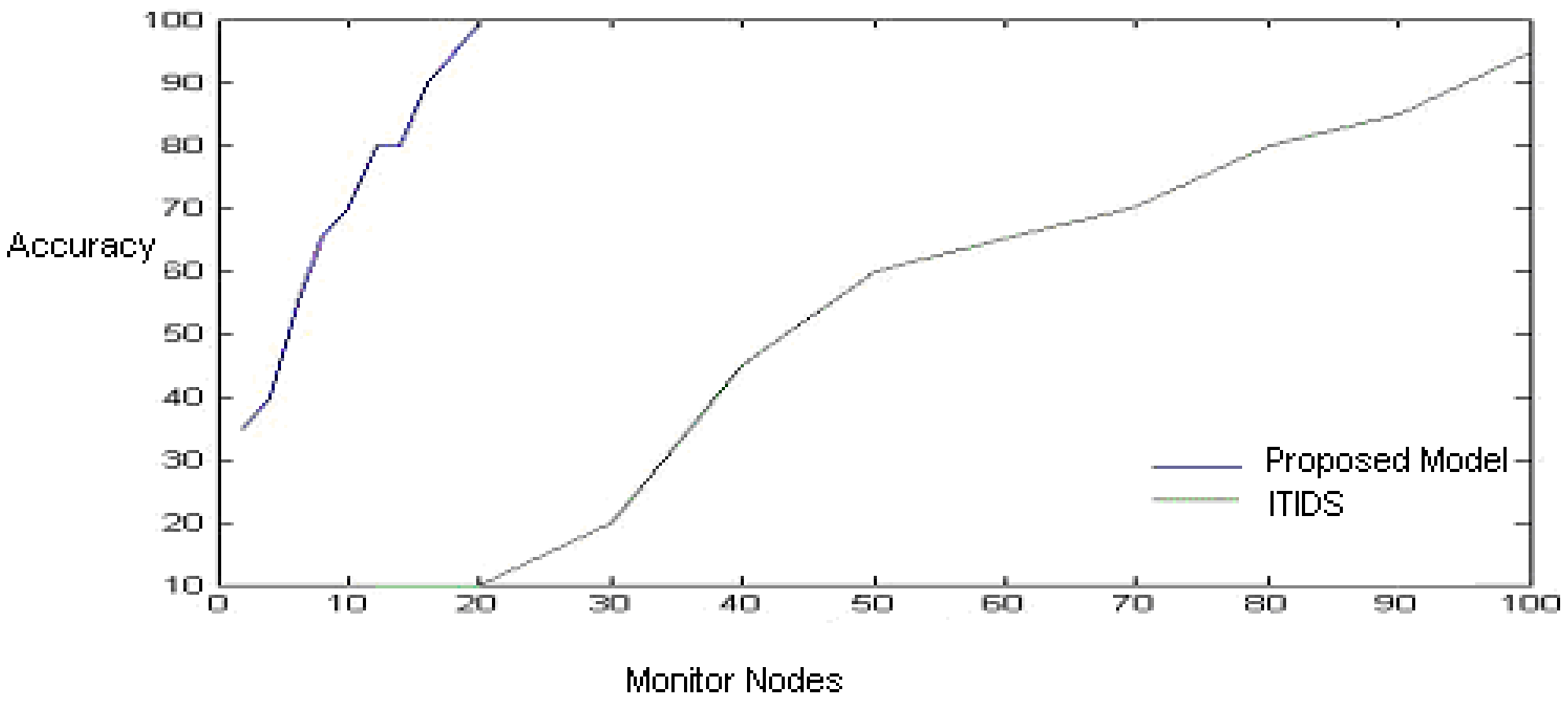

Figure 5 Comparison of Accuracy between ITIDS and Proposed Model

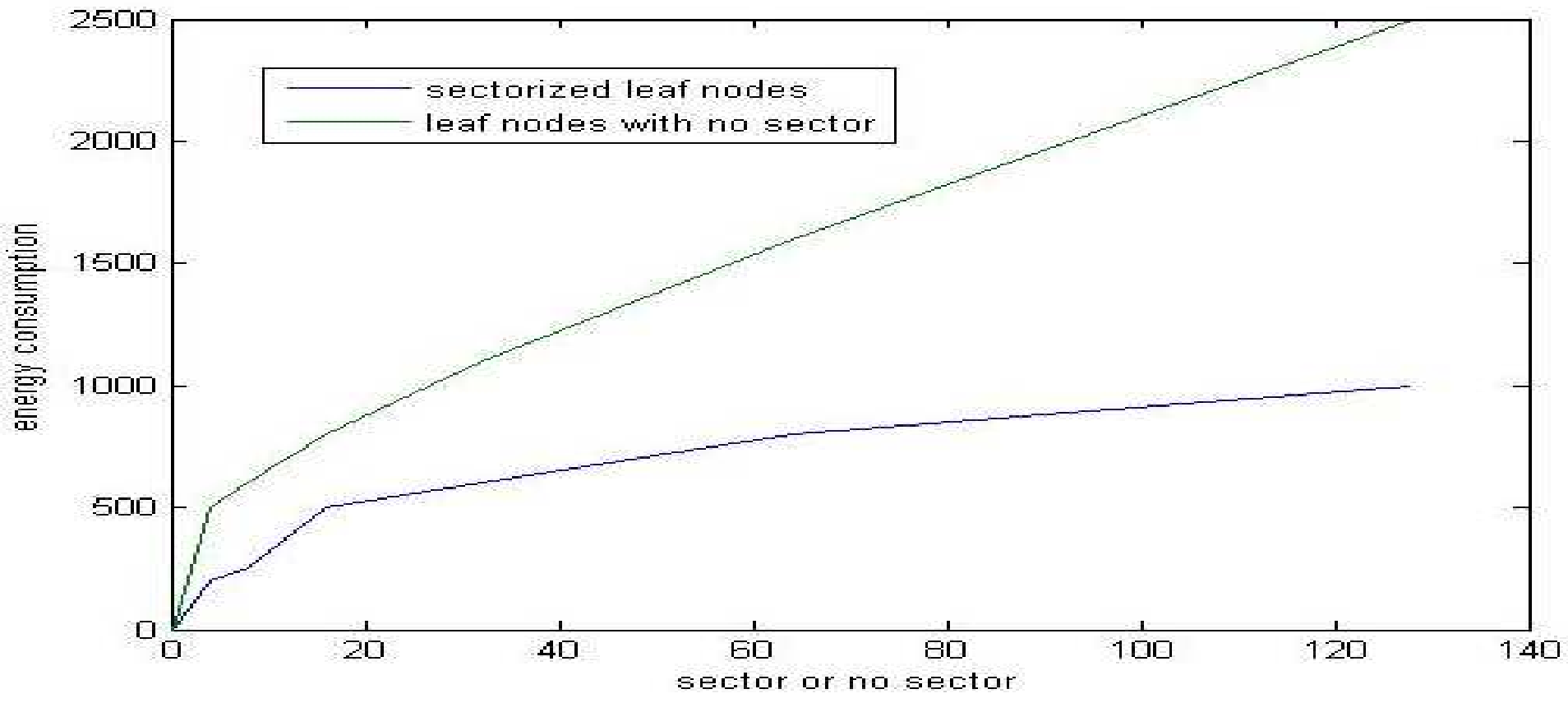

Figure 6 Energy Consumption with Sectorization and without Sectorization in Proposed Model

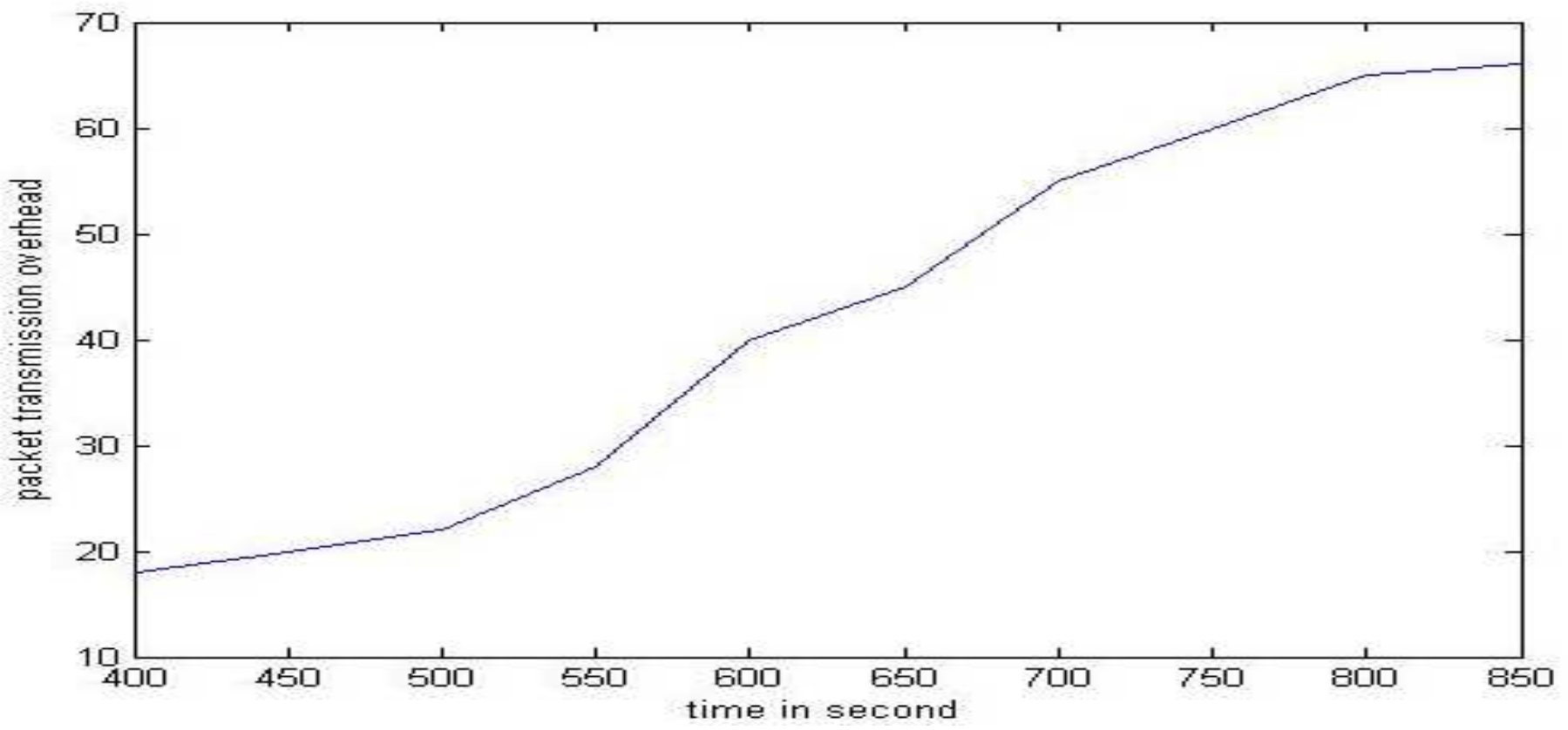

Figure 7 Packet Transmission Overhead with respect to Time in Second

## 6. Conclusion

Among the different types of prevalent attacks, sleep deprivation attack at link layer has been found to be the most devastating one for sensor nodes, exhausting the battery life very quickly. This paper comes up with the idea of a novel model that can detect sleep deprivation attack without using MAC based protocols like S-MAC, T-MAC, B-MAC, G-MAC. The outline of layer based approach using cluster technique to design a lightweight detection model capable of detecting insomnia of sensor nodes with less energy consumption has been documented here. The aim of this proposed model is to extend the lifetime of the HWSNET, even in the face of sleep deprivation attack. Generally, intruder attacks lower layer leaf nodes in HWSNET. In this article, insomnia detection is mainly focused on layer 1 that has no detection capacity of its own. More studies are being done to analyze the performance of the proposed model with respect to other considerable parameters such as energy consumption rate, delay, accuracy in presence of multiple intruders and can be compared with other existing models.

## References

[1] Mao, Y.,(2010), "A Semantic-based Intrusion Detection Framework for Wireless Sensor Network". In: 6th International Conference on Networked Computing (INC), Gyeongju, Korea, South.

[2] Misra, S., Venkata Krishna, P., Abraham, K.I., (2010), "Energy Efficient Learning Solution for Intrusion Detection in Wireless Sensor Networks". In: Proceedings of the 2nd international conference on Communication systems and Networks COMSNETS 2010

[3] Crosby, G.V., Hester, L., Pissinou, N., (2011), "Location-aware, Trust-based Detection and Isolation of Compromised Nodes in Wireless Sensor Networks". International Journal of Network Security , 107–117

[4] Chen, R.-C., Hsieh, C.-F., Huang, Y.-F., (March 2010) , "An Isolation Intrusion Detection System for Hierarchical Wireless Sensor Network". Journal of Networks

[5] Chen, R.-C., Huang, Y.-F., Hsieh, C.-F., (2010), "Ranger Intrusion Detection System for Wireless Sensor Networks with Sybil Attack Based on Ontology". New Aspects of Applied Informatics, Biomedical Electronics and Informatics and Communications

[6] Mamun, M.S.I., Sultanul Kabir, A.F.M., (July 2010), "Hierarchical Design Based Intrusion Detection System For Wireless Ad Hoc Sensor Network". International Journal of Network Security & Its Applications (IJNSA)

[7] Yan, K.Q., Wang, S.C., Liu, C.W., (2009), "A Hybrid Intrusion Detection System of Cluster-based Wireless Sensor Networks". In: Proceedings of the International MultiConference of Engineers and Computer Scientists, IMECS 2009, Hong Kong, March 18 - 20, vol. I

[8] Atakli, I.M., Hu, H., Chen, Y., Ku, W.-S., Su, Z., (2008) ,"Malicious Node Detection in Wireless Sensor Networks using Weighted Trust Evaluation". In: The Symposium on Simulation of Systems Security (SSSS 2008), Ottawa, Canada, April 14 –17

[9] Huo, G., Wang, X., (2008), "DIDS:A Dynamic Model of Intrusion Detection System in Wireless Sensor Networks". In: IEEE, International Conference on Information and Automation, Zhangjiajie, China, June 20 –23

[10] Stajano F. and Anderson R., (April 1999) ,"The resurrecting duckling: Security issues for ad-hoc wireless networks," in International Workshop on Security Protocols, pp. 172–194

[11] Techateerawat, P., Jennings, A., (2006), "Energy Efficiency of Intrusion Detection Systems in Wireless Sensor Networks". In: IEEE WIC 2006

[12] Bhattasali T. and Chaki R., (2011),"A Survey Of Recent Intrusion Detection Systems In Wireless Sensor Network". In: Proceedings of the Fourth International Conference on Network Security and Applications, CNSA 2011, Chennai, July 15~17

**Authors**

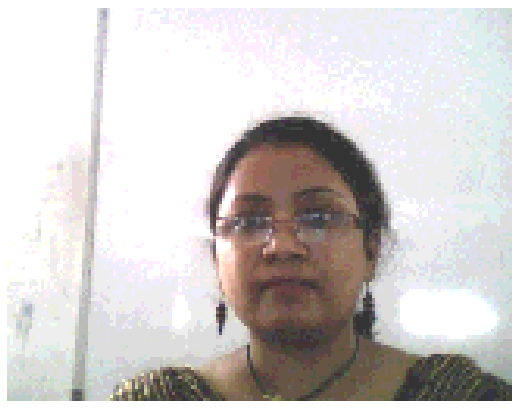

Tapalina Bhattasali has received her M. Tech. Degree in Information Technology from West Bengal University of Technology in 2011. She is at present working at Techno India College of Technology, Kolkata, India. Her research interests include the field of Wireless Sensor Network.

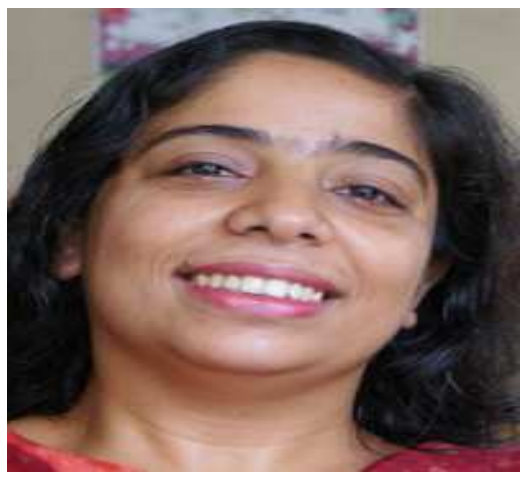

Rituparna Chaki is Associate Professor in the Department of Computer Science & Engineering of West Bengal University of Technology, Kolkata, India since 2005. She received her Ph.D. in 2002 from Jadavpur University, India. The primary areas of research interest for Dr. Chaki are Wireless Mobile Ad hoc Networks and Wireless Sensor Networks. She has also served as a Systems Manager for Joint Plant Committee, Government of India for several years before she switched to Academia. Dr. Chaki also serves as a visiting faculty member in other leading Universities including Jadavpur University. Dr. Chaki has number of international publications to her credit. Dr. Chaki has also served in the committees of several international conferences.